\documentstyle[12pt]{article}

\newcommand{\be}[3]{\begin{equation}  \label{#1#2#3}}
\newcommand{\ee}{ \end{equation}}
\newcommand{\ba}{\begin{array}}
\newcommand{\ea}{\end{array}}

\newcommand{\NP}[3]{{\em Nucl. Phys.}{ \bf B#1#2#3}}

\newcommand{\href}[1]{{\ref{#1}}}

\newcommand{\hcite}[1]{
 \cite{#1}}
\newcommand{\hbibitem}[1]{\bibitem{#1} 
  }
\newcommand{\hepth}[7]{
{hep-th/#1#2#3#4#5#6#7}}

\begin{document}


\thispagestyle{empty}
\rightline{HUB-EP-96/63}
\rightline{November 1996}
\rightline{hep-th/9611237}
\vspace{1truecm}
\centerline{\bf Decompactification near the horizon and non-vanishing
 entropy}
\vspace{1.2truecm}
\centerline{\bf Klaus Behrndt\footnote{Contribution to ``Strings '96'',
 Institute for Theoretical Physics, University of California at Santa
  Barbara, July 15--20, 1996. Supported by the DFG.}}
\vspace{.5truecm}
\centerline{Humboldt-Universit\"at, Institut f\"ur Physik}
\centerline{Invalidenstra\ss e 110, 10115 Berlin}
\centerline{Germany}
\vspace{1.2truecm}


\vspace{.5truecm}

\begin{abstract}
Intersecting $D$-brane configurations are related to
black holes in $D=4$. Using the standard way of compactification
only the Reissner-Nordstr{\o}m black hole is non-singular.
In this paper we argue, that also the other black holes are
non-singular if i) we compactify over a periodic array and ii)
we allow the string metric after reaching a critical curvature
to choose  the dual geometry. Effectively this means that near
the horizon the solution completely decompactifies and chooses
a non-singular $D$-brane configuration.
\end{abstract}

\newpage


\noindent
If the 4d spacetime exhibits singularities one usually argues that physics
breaks down there and if the singularity is not hidden by a horizon one
should throw away this solution. On the other side in string theory there
are some examples where these singularities are only a consequence of
a too restrictive field theoretical description. We assume that if the
string has the choice it always propagate in  a non-singular geometry. 
Mainly there are two ways

i) Decompactification 

ii) Transition into the dual geometry

\noindent
The first possibility is based on the fact singularities are improved
in higher dimensions. If the singularity disappears, it is not really
a physical singularity, but a singular compactification.  An example
is the rotating BPS black hole in 4 dimensions that has a naked
timelike singularity. In \hcite{ho/se} it has been argued that near
the core this solution decompactifies and the singularity becomes
null.  There are also other examples, where the solution becomes even
non-singular.  Below we will explain how this goes. The second
possibility is a true stringy feature. String theory exhibits many
additional symmetries that are not present in standard field
theories. These are the dualities, which allow the string to live in
different geometries. If the curvature reaches a certain level the
string theory undergoes a phase transition (from the momentum to
winding phase). From the string theory point of view both phases are
not distinguishable. They are completely equivalent even though the
singularity structure is different. This scenario has been discussed
for cosmological solutions by many authors, see e.g.\ \hcite{br/va},
\hcite{ga/ve} (pre-big-bang) or \hcite{lu/ov/wa} for a recent
discussion. In principle, there is yet a third possibility. Namely,
that the singularity disappears on the quantum level.  In this paper,
however, we want to restrict ourselves on pure classical scenario (see
e.g.\ \hcite{ho/ma}).

On the other side there are also arguments ``in favour'' of
singularities. Any modification of general relativity which is
completely non-singular cannot have a stable ground state
\hcite{ho/my}. Let us however discuss some examples for which we can
give a non-singular description. These are the 4d supersymmetric
static black holes and the non-singular geometry enables us to assign
them a non-vanishing entropy density.

\medskip

\noindent
Let us now explain how the two possibilities works and what are the
consequences for the black hole entropy.

\bigskip

\noindent
{ \bf 1) Decompactification}

\medskip

\noindent
Let us start with the example of the electric $a=1/\sqrt{3}$ black hole, 
which is given by
\be010
ds^2 = H^{-3/2} dt^2 - H^{3/2} d\vec{x}^2 \quad , \quad 
    e^{- 2 \phi /\sqrt{3} } =\sqrt{H} \quad , \quad  
F_{0m} \sim \partial_m H^{-1}\, ,
\ee
with the harmonic function $H(x)$. It can be obtained by compactification 
of the 5d Reissner-Nordstr{\o}m black-hole
\be020
ds^2 = \frac{1}{H^2} dt^2 - H (dx_5^2 + d\vec{x}^2) \quad , \quad 
F_{0m} \sim \partial_m H^{-1}\, ,
\ee
where $x_m = (x_5,\vec{x})$ and
\be030
H(x_5, \vec{x}) = 1 + \frac{r_h^2}{\rho^2}\, , 
\ee
with $\rho^2 = x_5^2 +
r^2$. After compactification of $x_5$ (with the dilaton related to
$g_{55}$) we get the black hole (\href{010}). For this, however,
we have to assume that $H$ is periodic: $x_5 \sim x_5 + 
2 \pi R$. Then we can make the standard ansatz for $H$ as a periodic array 
\hcite{ga/ha/li}
\be040
H = 1 + \sum_{n= - \infty}^{+\infty} \frac{r_h^2}{r^2 +
   (x_5 + 2 \pi n R)^2} = 1 + \frac{r_h^2}{2R \, r}
   + {\cal O}(e^{-\frac{r}{R}}) \ .
\ee
Thus, away from the origin the dependence on $x_5$ is exponentially
suppressed, this direction is compactified on a circle with radius
$R$.  On the other side near the horizon ($\rho=0$) the
$5th$ coordinate becomes visible and the solution decompactifies to its 5d
origin. The asymptotic behaviour of the metric
near the horizon is given by
\be050
ds^2 \rightarrow \left( \frac{\rho^2}{r_h^2} \right)^2 \, dt^2 - 
r_h^2 \left( \frac{d\rho}{\rho} \right)^2 - 
 r_h^2 d\Omega_3^2 = e^{4 \eta / r_h} dt^2 - d\eta^2 - r_h^2 d\Omega_3^2\, ,
\ee
where $\rho/r_h = e^{\eta / r_h}$ and $r_h$ is the radius of the $S_3$ sphere.
Hence, the asymptotic geometry is: $(\mbox{de Sitter})_2 \times S_3$ which
is non-singular. 

On the other side if we assume that the harmonic function $H$ depends only
on $\vec{x}$ and not on $x_{5}$ the scalar field is again divergent
indicating that the compactification radius becomes infinite. Now, however,
the radius of the $S_{3}$ does not stay finite. Instead it shrinks to zero
yielding a singularity. This singularity is a consequence that we have
been too restrictive in the coordinate dependencies.


\bigskip

\noindent
{ \bf 2) Dual geometries }

\medskip

\noindent
We explained already at the beginning that special symmetries allows
the string to choose different geometries. To be explicit let us
consider an euclidian model of a spatially flat space which is
compactified on a torus with radii given by $e^{2 c \eta}$ for some
constants $c$.  This space is the near-horizon
geometry of extremal $p$-branes and the metric is given by
\be060
ds^{2} = d\eta^{2} + e^{2 c \eta} d\vec{z}^{\,2}
\ee
where $\vec{z}$ are some spatial coordinates (world-volume coordinates
for the $p$-brane). In the cosmological setting the big-bang is given
by $\eta \rightarrow -\infty $, where $e^{2 c \eta}$ shrinks to
zero. But what happens if we approach this point?  We assumed that the
spatial coordinates $\vec{z}$ are compactified with the radius $e^{c
\eta}$. For any compactified string theory there are two different
kind of modes. One are the standard momentum modes which rule in the
regime of large compactification radii. The other are the winding
modes that become important for small radii. If the compactification
radius is of order one ($\eta=0$) a phase transition happen, the
momentum modes drop out and the winding modes become important. The
consequence is that the string feels a different geometry where the
radii are inverted
\be070
\tilde{ds}^{2} = d\eta^{2} + e^{- 2 a \eta} d\vec{z}^{2} 
\ee
Going now to negative values of $\eta$ the spatial geometry now starts to
expand, the compactification radii increase. This is the pre-big-bang
region \hcite{ga/ve}. This scenario has been discussed for many cases
\hcite{br/va}. Some of these models have a direct conformal field
theory description \hcite{lu} which makes the non-singular nature
obvious. This effect is a consequence that in string theory, below a certain
length scale, the usual local field theory loses their sense.  There is
a lower bound for space time structure that can be resolved and many
singularities are hidden. Expressed in terms of thermodynamics it
means that that there is a upper bound for the temperature ($T \sim 1/L$, 
where $L$ is the compactification radius) and the maximum
temperature is known as the Hagedorn temperature \hcite{br/va}.


\bigskip

\noindent
{ \bf 3) Consequences for black hole entropy}

\medskip

\noindent
The supersymmetric black holes can be seen as compactified
intersection of branes. In this paragraph we discuss what happens with
the black holes when we approaching the horizon.  Let us start with a
short survey of the known supersymmetric black holes.

A common way to classify four--dimensional Maxwell/scalar black hole
(BH) solutions is to specify the coupling of the scalar fields to the
gauge fields.  In the simplest case of only one scalar field and one
gauge field this coupling is characterized by a single parameter $a$
and the action in the Einstein frame is given by 
\be080 
 S_{4d} = \frac{1}{16 \pi G_4} \int d^4 x \sqrt{|g|} 
 \{-R + 2 (\partial \phi)^2 + e^{-2 a \phi} F^2 \} \ , 
\ee
where $G_4$ is the 4--dimensional Newton constant.  There are four
different types of extremal\footnote{In this letter we consider only
extremal solutions.} black hole solutions, which are defined in terms
of a function $H$ which is harmonic on the 3--dimensional
transverse space. The metric of these solutions is given by
\be090
\ba{llll}
a=0 & : & ds^2 = H^{-2} dt^2 - H^2 d\vec{x}^2 \quad , & e^{-2 \phi} =1 \ , \\
a=1/\sqrt{3} & : & ds^2 = H^{-3/2} dt^2 - H^{3/2} d\vec{x}^2 \quad , 
    & e^{\pm 2 \phi /\sqrt{3} } =\sqrt{H}  \ , \\
a=1 & : & ds^2 = H^{-1} dt^2 - H d\vec{x}^2 \quad , & e^{\pm 2 \phi} =H  \ ,\\
a=\sqrt{3} & : & ds^2 = H^{-1/2} dt^2 - H^{1/2} d\vec{x}^2 \quad , 
 & e^{\pm 2 \phi/\sqrt{3}} = \sqrt{H}  \ .
\ea
\ee
These solutions have been generalized to different harmonic functions
in \cite{ra} (for $a=0$ this generalization has been given in
\cite{cv/ts}).  For a discussion of these solutions as bound states,
see \cite{du/ra}.  The four solutions (\href{090}) are also known as
\medskip

\begin{tabular}{lll}
$a=0$ & :  & 4d Reissner-Nordstr{\o}m (RN) solution, \\
$a=1/\sqrt{3}$ & : & 5d RN  ($\phi$ is a modulus field), \\
$a=1$ & :  & dilaton black hole ($\phi$ has standard dilaton coupling), \\
$a=\sqrt{3}$ & : & 5d KK black hole ($\phi$ is a modulus field). 
\end{tabular}   \smallskip
\medskip

\noindent
For $a\ne 0$ the gauge fields can be electric or magnetic. The two
possibilities correspond to different signs of the scalar field
$\phi$. In formula (\href{090}) the ``+'' sign corresponds to the
magnetic case. On the other hand, the $a=0$ RN solution is dyonic.  It
turns out that in four dimensions only the $a=0$ RN solution is
non-singular at the horizon $r=0$. However, following \cite{gi/ho/to},
also the $a\ne 0$ solutions can be understood in a non-singular way,
in the sense that they follow from the dimensional reduction of the
following higher--dimensional non-singular solutions \hcite{be/be}
\medskip

\begin{tabular}{lll}  
$a=1/\sqrt{3}$ & : & 5d RN electric black hole or magnetic string\, , \\
$a=1$ & : & 6d self-dual string\, , \\
$a=\sqrt{3}$ & : & 10d self-dual $D$-3-brane\, . 
\end{tabular}   \smallskip
\medskip

\noindent
In addition, these are just the brane solutions that exhibits supersymmetry
restoration near the horizon (see \hcite{fe/gi/ka} and refs.\ therein).
So, the singularities in the solution (\href{090}) can be seen as a
consequence of the compactification. For the case of the electric
$a=1/\sqrt{3}$ solution (\href{010}) we have seen already that it
decompactified into a non-singular solution (\href{020}) when we allow
the harmonic functions to depend also on $x^5$.  A singularity appears
only if we are to restrictive in the compactification. On the other
side for the magnetic $a=1/\sqrt{3}$ case things work differently.
Here, the non-singular analogue is the magnetic string in 5 dimensions
where we have wrapped the string around the $5th$ direction. Now, the
transversal space stays 3-dimensional but if we approach $r=0$ the
radius of the $5th$ coordinate shrinks and would vanish at $r=0$. As we
argued in the last paragraph this situation will not happen in string
theory. Instead, at a certain point a phase transition will happen
(the momentum modes drop out and the winding modes become important).
For the geometry this means that the compactification radius inverts
and expand if we further approach the point $r=0$ where finally the
string is completely decompactified. For the other black holes ($a=1$
and $a=\sqrt{3}$) we argue similar, but now both effects happen at the
same time. The $a=1$ black hole decompactifies completely to the 6d
self-dual string and the $a=\sqrt{3}$ solution to the 10d self-dual
3-brane. The decompactification of the transversal space can
really be seen in our field theoretical formulation. On the other
side the inversion of the compactification radius is a dynamical
process and related to a phase transition at the Hagedorn temperature.
Even more, the inversion of the world volume radii makes them
to transversal coordinates, i.e. the magnetic strings becomes a black
hole. Basically, in string theory what we have to expect is that
the object in the target space (e.g.\ 2-brane or 1-brane) is not
well defined. The string theory chooses that object which is least
singular. The confusion here comes from the point that we are looking
on the object from two different sides, from the momentum modes and
from the winding modes. But the main point is that the 
compactification radius becomes effectively infinite. 

What is this meaning for the entropy? There are two points of view.
First, we should only consider the transversal space and define an
entropy density (entropy per unit world-volume).  In this picture we
only have to take into account possible decompactifications of the
transversal space, when we approach the horizon. We do not get into
trouble with the phase transition. This procedure is well defined in
the field theory. A more subtle point is the total entropy, where one
integrates also over the world-volume coordinates. Here we do not have
a standard field theory description, or if we simply integrate we get
zero entropy (the radii vanish).  This however is only a consequence
that we have throw away possible winding modes.  For vanishing radii
the momentum modes are completely suppressed but the winding states
have a high degeneracy. To count the degeneracy of winding states is
its own problem, but assuming that the number of momentum modes for
large compactification radius coincides with the number of winding
modes at small radii we can simply count the states by going on
the brane and simultaneously making the compactification radius
infinite ($L \rightarrow \infty$, where $L$ is the periodicity of the
coordinate). Thus we have the situation of ``$ 0 \cdot \infty$''
(infinite world volume but zero metric components), which could
give a finite result for the entropy. A further way to get a final
total entropy is to calculate the expression not {\em on} the
horizon, but approach the horizon as far as possible. This means
we calculate the entropy just at the critical point, $\eta =0$ in 
(\href{060}), which is the $T$-self-dual radius. 
This defines us the stretched horizon \hcite{su}.

To avoid this subtleties we discuss the entropy density for the black
holes in (\href{090}). First, we note that the radius of all horizons
is given by $r_h$, which can be expressed by the electric and magnetic
charges. Since the 4 dimensional solution can be electric as well as
magnetic, let us distinguish between both charges. Then, for $a=0, 1,
\sqrt{3}$, after decompactification of the transverse space and
after integrating over the different spherical parts $S_k$
(i.e.~$k = 2+a^2$) the entropy per unit world volume can be written as
\hcite{be/be}
\be100
\ba{rl}
& S = \frac{A}{4 \, G_k} = \frac{1}{4 \, G_k} \int_{S_k}
(r_h)^k d\Omega_k = \pi (r_h)^k \\[2mm]
\mbox{with:}& (r_h)^2 = 
   4  \sqrt{\left(( \vec{n} + \frac{1}{2} \vec{Q}) L (\vec{n} + 
   \frac{1}{2} \vec{Q})\right) \left( (\vec{p} + \frac{1}{2} \vec{P}) 
   L (\vec{p} + \frac{1}{2} \vec{P})\right)} \ .
\ea
\ee
where $L$ is the metric in the $O(d,d)$ space\footnote{In this formula
we have already included the different ways of embedding the
intersection into the 10d space, which causes this $O(d,d)$
structure.}  and $\vec{n}$ and $\vec{p}$ are arbitrary unit vectors
($\vec{n} L \vec{n} = \vec{p} L \vec{p} =1$).  With $G_k$ we take into
account that the Newton constant has to be rescaled when one compares
expressions in different dimensions.  In our normalization in 4
dimensions we have $G_4=1$. This is a generalization of the entropy
formula of Larsen and Wilczek \hcite{la/wi}. It reduces to their
expression in the limit $\vec{n} \cdot \vec{Q} = \vec{p} \cdot
\vec{P} = 0$.

In our approach the case $a=1/\sqrt{3}$ is special. The electric case
yields, integrating over $S_3$, an entropy density $S\sim \sqrt{Q^3}$
whereas the magnetic case leads, after integrating over $S_2$, to
$S\sim P^2$.  Many authors have investigated the electric case (see
e.g.~\cite{st/va}). However, it does not fit into our entropy formula
(we have 3 intersecting branes in this case).  Note that the
$a=1/\sqrt 3$ BH is also special in the sense that it cannot be
expressed by $D$-3-branes only, which are the only non-singular branes
in 10 dimensions. In order to get a non-singular result we need to
include a boost or Taub-NUT soliton, i.e.\ KK modes. Thus, we can
conclude that the formula (\href{100}) describes the entropy for all
BH's that can be expressed in a non-singular way by $D$-3-branes only.


\newpage


\end{document}